\documentclass[amsmath,11pt]{article}
\usepackage{amsfonts,amsmath,amssymb,graphics,epsfig}
\usepackage{caption}\usepackage{subcaption}
\usepackage{enumerate}\usepackage{theorem}
\usepackage{xcolor}\usepackage{ulem}

\def\d {\mbox{d}}

\newcommand{\PC}[1]{\ensuremath{\left(#1\right)}}

\begin{document}

\title{\bf Timelike vs null deceleration parameter\\ in tilted Friedmann universes}

\author{Jessica Santiago${}^{1}$ and Christos G. Tsagas${}^{1,2}$ \\${}^1${\small Section of Astrophysics, Astronomy and Mechanics, Department of Physics}\\ {\small Aristotle University of Thessaloniki, Thessaloniki 54124, Greece}\\ ${}^2${\small Clare Hall, University of Cambridge, Herschel Road, Cambridge CB3 9AL, UK.}}

\date{\empty}

\maketitle

\begin{abstract}
Information exchanged between observers in the universe typically travels along the null rays of the associated light signals. One may therefore decompose the luminosity distance of a given radiation source along these null geodesics, instead of the timelike worldlines of the emitting/receiving observers. In so doing, one obtains (among others) the null counterparts of the familiar Hubble and deceleration parameters. Although the timelike and the null sets of these parameters coincide in an exact Friedmann universe, they generally differ in perturbed cosmological models. The situation becomes more involved in tilted cosmologies, namely in those equipped with two families of relatively moving observers. There, the two observer groups generally disagree on the values of these parameters, simply because of their relative motion. Assuming a tilted, perturbed Friedmann universe with dust, we find that the relative-motion effects alone can locally change the sign of the null deceleration parameter from positive to negative, while at the same time the host universe is globally decelerating. An exactly analogous effect has also been reported with regard to the timelike deceleration parameter. The difference is that, in the null case, the impact of the relative motion is twice as strong, thus making it easier for the unsuspecting observers to misinterpret a local change in the sign of the deceleration parameter as recent global acceleration.
\end{abstract}

\section{Introduction}\label{sI}
Observers moving with respect to each other generally experience different versions of ``reality'', entirely because of their relative motion. This is particularly true in relativity, where time and space are no longer absolute and observer-independent physical entities. Applied to cosmology, this principle implies that observers living in typical galaxies, which move relative to each other -- as well as relative to the universal expansion, may also disagree on the specifics of their host universe. Then, the extent of the disagreement should in principle depend on the speed of the observers' relative motion, as well as on the size of the moving domain they happen to live in.

Over the years, observations have repeatedly verified the presence of large-scale peculiar motions in the universe, namely large sections of the observable cosmos moving coherently towards a given direction in the sky (e.g.~see~\cite{Aetal} and references therein). These are the so called ``bulk flows'', with typical sizes from few hundred to several hundred Mpc and typical velocities ranging between few hundred and several hundred km/sec. Peculiar velocities are defined and measured with respect to the reference coordinate system of the Cosmic Microwave Background (CMB), where the associated dipole vanishes by default. Our Local Group of galaxies, for example, drifts with peculiar velocity slightly faster than 600~km/sec (e.g.~see~\cite{Koetal,Agetal}).\footnote{The CMB frame is also referred to as the Hubble frame, so we will occasionally use these two terms interchangeably.} Overall, it is fair to say that no real observer in the universe follows the idealised CMB frame, but we all have some finite peculiar velocity relative to it.

Given a family of observers, their motion is monitored by the kinematics of their associated timelike worldlines. These determine whether the observers move apart or come closer to each other, whether they rotate and are subjected to shearing forces, as well as whether they live along geodesics or not. In cosmology, the expansion rate of an observer group (essentially the value of the Hubble parameter measured by them) follows from the divergence of their 4-velocity field. The time derivative of the latter determines the acceleration/deceleration of the expansion, while the higher-order temporal derivatives are related to additional parameters that are typically referred to as ``jerk'', ``snap'', etc. (see~\cite{V} for a comprehensive presentation and discussion). All these kinematic features emerge naturally after performing a Taylor-series expansion to the luminosity distance of a given source, expressed in terms of redshift along the observers' timelike worldlines.

On the other hand, information exchange in the universe typically propagates via electromagnetic or gravitational-wave signals, both of which travel along null geodesics. One may therefore choose to monitor the universal kinematics, for instance its expansion rate and its acceleration/deceleration rate along the observer's line of sight~\cite{CM}-\cite{H}. More specifically, following in the steps of~\cite{V}, one can Taylor expand the luminosity distance of a source along the null geodesics of the emitted/received light rays and in so doing obtain the ``null'' analogues of the familiar ``timelike'' parameters, like the Hubble, the deceleration, the jerk and the snap parameters, as well as of their higher-order counterparts~\cite{H}. Not surprisingly, the aforementioned timelike and null parameters are related to each other. In fact, assuming a spatially homogeneous and isotropic -- that is a Friedman-Lemaitre-Robertson-Walker (FLRW) -- universe, the aforementioned two sets coincide. In general, however, there are differences even at the linear perturbative level (i.e.~in almost-FLRW cosmologies).

The situation becomes more involved when the perturbed universe is also ``tilted'', which provides a more realistic theoretical model that can accommodate two (or more) families of relatively moving observers (e.g.~see~\cite{KE,HUW}). Typically, one of these groups follows the CMB frame, which also defines the reference coordinate system of the universe, while the other moves relative to it and lives in a typical galaxy like our Milky Way. Even at the linear level, the cosmological parameters measured in the two frames generally differ, with their difference caused by the observers' peculiar motion alone. The strength of the effect depends on the magnitude of the peculiar velocities and on the size of the bulk-flow domain. More specifically, the faster the drift velocity the stronger the impact of the effect, while the larger the size of the bulk-flow the wider the range of the effect.

The impact of relative motion on the timelike deceleration parameter was originally investigated in~\cite{T1,T2} and subsequently refined in~\cite{TK,T3}. Using linear relativistic cosmological perturbation theory, it was shown that observers living inside bulk flows, like those reported in various surveys (e.g.~see~\cite{ND}-\cite{Scetal} for a representative though incomplete list), could assign negative values to their local deceleration parameter, while at the same time the host universe was globally decelerating. Therefore, the universal acceleration ``experienced'' by these bulk-flow (tilted) observers is a mere local illusion, solely triggered by their relative motion. In the aforementioned theoretical studies, the analysis took place in a perturbed Einstein-de Sitter universe, primarily for mathematical simplicity, since the results hold on essentially all FLRW universes, irrespective of their spatial curvature and equation of state~\cite{T4}. Not surprisingly, the relative-motion effects were found to increase closer to the observer and to fade away on progressively longer distances. In particular, the local deceleration parameter was found to become increasingly less negative away from the bulk-flow observer, to turn positive at a certain threshold and approach its CMB value at sufficiently high redshifts, thus mimicking the phenomenology of a recently accelerating universe. The ``transition length'', namely the scale where the sign of the local deceleration parameter changed from positive to negative was first established in~\cite{T3}, where it was also estimated to vary between few hundred and several hundred Mpc, depending on the reported size and speed of the bulk flow in question.

Although observations are used to measure the timelike deceleration parameter, here we will turn our attention to its null counterpart and consider the effects of the large-scale peculiar flows on its interpretation. Recall that this is the deceleration parameter measured along the null geodesics of the travelling radiation signals, exchanged between (idealised) observers following the CMB frame, or between (real) observers living inside a bulk flow and moving along the tilted frame. In so doing, we employ the 1+3 covariant approach to relativity and cosmology (see~\cite{Eh,El} and also~\cite{TCM,EMM} for recent reviews), which allows for a direct and transparent comparison between the relative motion effects. We pose the following questions: \textit{(i) Do both timelike and null deceleration parameters exhibit the same type of behaviour in the presence of peculiar motions? (ii) How and to what extent peculiar velocities affect each one of these parameters?} In what follows, we will look for answers by employing linear relativistic cosmological perturbation theory. In practice, this means that our analysis applies to scales close and beyond the 100~Mpc mark, which typically sets the nonlinear threshold.

The starting point are the nonlinear ``timelike'' and ``null'' kinematic equations. The former are given in~\cite{TCM,EMM}, while the latter can be found in~\cite{CM}-\cite{H}. Adopting the notation of~\cite{H} for our null formulae,, we linearise both sets around an FLRW universe, assuming low-energy matter with zero pressure. This allows us to establish the linear relation between the null and the timelike deceleration parameters and also verify that they coincide in an exact Friedmann universe. We then proceed to introduce a tilted almost-FLRW cosmology equipped with two groups of observers, namely those aligned with the CMB frame and those moving relative to it. We find that the null deceleration parameter measured by the fictitious CMB observers and the one measured in the tilted frame of their real counterparts generally differ. The difference is entirely due to the relative motion of the two frames and depends on the speed, as well as the scale of the bulk flow in question. These results are in full qualitative agreement with those reported on their timelike analogues (e.g.~see~\cite{TK}-\cite{T3}). It should also be noted that we come to these conclusions after taking the viewpoint of both the CMB observers and of their bulk-flow counterparts.

Quantitatively speaking, our linear analysis shows that the relative-motion effects on the null deceleration parameter are twice stronger than those on its timelike analogue. Therefore, peculiar motions can make the local null deceleration parameter appear even more negative than its timelike counterpart. In addition, the transition length, namely the scale where the sign of the null deceleration parameter changes from positive to negative, is larger than that of its tilmelike analogue. Overall, the relative-motion effects are more pronounced on the null deceleration parameter than on its timelike counterpart. As in the timelike studies, the negative values of the deceleration parameter measured in the tilted frame of the real observers are simply a local artefact of their peculiar motion. Indeed, as shown in~\cite{T1}-\cite{T3} as well as in \S~\ref{sRMEQ} here, the host universe is still globally decelerating. To the unsuspecting observers, however, an apparent local change in the sign of the decelaration parameter may be easily misinterpreted as recent global acceleration.

\section{Two sets of deceleration parameters}\label{sTSDPs}
In cosmology, the familiar deceleration parameter ($q$) is measured along the timelike worldlines of a family of observers. Alternatively, it is possible to introduce a deceleration parameter ($\mathfrak{Q}$) defined along the null geodesics of electromagnetic signals traveling through the universe. Hereafter, we will respectively refer to $q$ and $\mathfrak{Q}$ as the \textit{timelike} and the \textit{null} deceleration parameters.

\subsection{The ``timelike'' deceleration parameter}\label{ssTDP}
Consider a family of observers living along timelike worldlines (not necessarily geodesics) tangent to the 4-velocity field $u_a$, normalised so that $u_au^a=-1$. This 4-vector defines the temporal direction, while the associated 3-space is determined by the projection tensor $h_{ab}=g_{ab}+u_au_b$, with $g_{ab}$ representing the spacetime metric. The projector acts as the metric of the observers' 3-D rest space (strictly speaking when there is no rotation) and also satisfies the constraints $h_{ab}u^a=0$, $h_{ac}h^c{}_b=h_{ab}$ and $h_a{}^a=3$ by construction. In addition, together with the $u_a$-field, the projection tensor introduces an 1+3 splitting of the spacetime into time and 3-D space. Then, assuming that $\nabla_a$ is the 4-D covariant derivative operator,  overdots denote differentiation with respect to time (i.e.~${}^{\cdot}=u^a\nabla_a$), while ${\rm D}_a=h_a{}^b\nabla_b$ represents the 3-D (spatial) covariant derivative operator (e.g.~see~\cite{TCM,EMM}).

All the information regarding the kinematics of the aforementioned group of observers is encoded in the gradient $\nabla_bu_a$. One may decode the data by introducing the decomposition
\begin{equation}
\nabla_bu_a= {1\over3}\,\Theta h_{ab}+ \sigma_{ab}+ \omega_{ab}- A_au_b\,,  \label{Nbua}
\end{equation}
which brings forward the motion's irreducuble kinematic variables~\cite{TCM,EMM}. These are the volume scalar ($\Theta={\rm D}^au_a$), the shear and the vorticity tensors  ($\sigma_{ab}={\rm D}_{\langle b}u_{a\rangle}$ and $\omega_{ab}={\rm D}_{[b}u_{a]}$ respectively, with $\sigma_{ab}u^b=0=\omega_{ab}u^b$) and the 4-acceleration vector ($A_a=\dot{u}_a$, with $A_au^a=0$). The former indicates expansion/contraction when positive/negative respectively, while nonzero shear and vorticity ensure the presence of kinematic anisotropies and rotation. Finally, the 4-acceleration implies that there are non-gravitational forces in action and that the observes' timelike worldlines are not geodesics.

Assuming an expanding universe, the rate of the expansion is determined by the magnitude of the Hubble parameter ($H$), with the associated deceleration/acceleration monitored by the deceleration parameter ($q$). These are related to the volume scalar ($\Theta$) by means of
\begin{equation}
H= {\Theta\over3} \hspace{15mm} {\rm and} \hspace{15mm} q= -\left(1+{\dot{H}\over H^2}\right)= -\left(1+{3\dot{\Theta}\over\Theta^2}\right)\,,  \label{H-q}
\end{equation}
where overdots indicate proper-time derivatives (e.g.~$\dot{\Theta}= {\rm d}\Theta/{\rm d}\tau$). By construction, positive values for $q$ imply decelerated expansion, whereas negative ones ensure cosmic acceleration.

\subsection{The ``null'' deceleration parameter}\label{ssNDP}
Given a class of observers with 4-velocity $u_{a}$, one can define the unit basis $\{{e}_1, {e}_2, {e}_3\}$ to span the 3-D hypersurface orthogonal to the $u_a$-field (i.e.~$u_ae^a=0$, $h_a{}^be_b=e_a$ and $e_ae^a=1$). We may therefore assume that $e_a$ is the spatial direction along which the observers collect their incoming data. Then, the photon 4-momentum ($k_{a}$) can be decomposed in terms of $u_{a}$ and ${e}_a$ as
\begin{equation}
	\label{k}
	k_a = E(u_a -e_a)\,.
\end{equation}
Note that $E=-k_au^a$ is the energy density of photons traveling along null geodesics (tangent to the $k_a$-vector, with $k_ak^a=0$) and measured along the direction of the spacelike $e_a$-vector by observers living along timelike worldlines tangent to the $u_a$-field.

The luminosity distance of a given radiation source can be expanded into a Taylor series in terms of redshift. When the expansion takes place along the observers' timelike worldlines, the coefficients are expressed in terms of the familiar Hubble ($H$) and deceleration ($q$) parameters, as well as in terms of higher-order derivatives of the cosmological scale factor (see~\cite{V} for further discussion). An analogous series expansion along the null geodesics of the electromagnetic signals, leads (among others) to the effective Hubble ($\mathfrak{H}$) and deceleration ($\mathfrak{Q}$) parameters associated with the null congruence~\cite{H}. These are respectively given by
\begin{equation}
	\label{fancyh}
	\mathfrak{H}\equiv -\frac{1}{E^2}\frac{{\rm d}E}{{\rm d}s} \hspace{15mm} {\rm and} \hspace{15mm}
	\mathfrak{Q}\equiv -1 - \frac{1}{E\mathfrak{H}^2}\frac{{\rm d}\mathfrak{H}}{{\rm d}s}\,,
\end{equation}
where $s$ is the null affine parameter and ${\rm d}E/{\rm d}s<0$ in an expanding universe. Comparing the above definitions of $\mathfrak{H}$ and $\mathfrak{Q}$ to those of their standard (timelike) counterparts, shows that the photon energy density ($E$) acts as an effective cosmological scale factor ($a$), while $s$ plays the role of the conformal time ($\eta$). Recall that $H=({\rm d}a/{\rm d}\eta)/a^2$ and $q=-1-({\rm d}H/{\rm d}\eta)/aH^2$, with $\dot{a}/a=H$ and $\dot{\eta}=1/a$.\footnote{It is also worth noticing that the energy of a photon traveling in a FLRW metric is given by $E = E_0/a$. Differentiating both sides with respect to the photon's affine parameter we obtain:
\begin{eqnarray}
-\frac{1}{E^2}\frac{{\rm d}E}{{\rm d}s} =  \frac{1}{E_0} \frac{\d a}{\d s}.
\end{eqnarray}
Now, using equation \eqref{k} to rewrite $(\d /\d s)$ in terms of $(\d /\d \eta)$, we have:
\begin{eqnarray}
\mathfrak{H} = \frac{1}{E_0}\PC{E\; \frac{\d a}{\d \eta} -E \; e^{\mu}\partial_{\mu}a } =  H - \frac{1}{a} \;e^{\mu}\partial_{\mu}a \; .
\end{eqnarray}
Clearly, for an FLRW metric, $a = a(\eta)$, ensuring that the last term on the right-hand side is zero and $\mathfrak{H} = H$. However, allowing for the presence of spatial inhomogeneities, these terms might differ. Notice also that the evolution equation for a photon traveling along a null geodesic in a FLRW metric is given by:
\begin{equation}
\label{EFRW}
-\frac{1}{E^2}\frac{d E}{ds} = H.
\end{equation}
Comparing equations (\ref{fancyh}a) and \eqref{EFRW} becomes clear that the extension of the expansion parameter to the null case is very straightforward and has a physically meaningful interpretation.		
The only caution note that must be kept in mind, however, is that talking about a scale factor beyond the FLRW metrics context might become an awkward and cumbersome subject. We therefore keep this merely as a curiosity note.}

The effective $\mathfrak{H}$ and $\mathfrak{Q}$ parameters, measured along the spatial direction $e_a$, can be decomposed into scalar, vector and higher-rank tensor parts, which respectively correspond to monopole, dipole and higher-order moments. These can then be expressed in terms of the kinematic variables of the associated observers. More specifically, the null Hubble parameter satisfies the expression~\cite{Cetal,H}
\begin{equation}
\mathfrak{H}(e)= {1\over3}\,\Theta- A_ae^a+ \sigma_{ab}e^ae^b\,.  \label{nH}
\end{equation}
Recall that $\Theta$, which is the volume scalar of the observers' timelike worldlines, is related to the familiar Hubble parameter ($H$) by means of $\Theta=3H$. According to the above, in an exact (unperturbed) Friedmann spacetime where $A_a=0=\sigma_{ab}$ by default, the null Hubble parameter has only a monopole (isotropic) component and coincides with its (standard) timelike counterpart (i.e.~$\mathfrak{H}=H$).

An analogous decomposition also applies to the null deceleration parameter, which when $\mathfrak{Q}$ is measured along the direction of the spatial $e_a$-vector, reads
\begin{equation}
	\label{fancyQcomplete}
\mathfrak{Q}(e)= -1- {1\over\mathfrak{H}^2(e)} \left(\overset{0}{\mathfrak{q}}+{\overset{1}{\mathfrak{q}}}_ae^a +\overset{2}{\mathfrak{q}}_{ab}e^ae^b +{\overset{3}{\mathfrak{q}}}_{abc}e^ae^be^c +\overset{4}{\mathfrak{q}}_{abcd}e^ae^be^ce^d\right)\,,
\end{equation}
where
\begin{eqnarray}
\overset{0}{\mathfrak{q}}&=& \frac{1}{3}\,\dot{\Theta}+ \frac{1}{3}\,{\rm D}_aA^a- \frac{2}{3}\,A_aA^a- \frac{8}{15}\,\sigma_{ab}\sigma^{ab}\,,  \label{q0}\\
\overset{1}{\mathfrak{q}}_a&=& -\dot{A}_{\langle a\rangle}- \frac{1}{3}\,{\rm D}_a\Theta- \frac{2}{5}\,{\rm D}^b\sigma_{ab}+ \omega_{ab}A^b+ 3\sigma_{ab}A^b- \frac{6}{5}\,\sigma_{ab}A^b\,, \label{q1}\\ \overset{2}{\mathfrak{q}}_{ab}&=& \dot{\sigma}_{\langle ab\rangle}+ {\rm D}_{\langle a}A_{b\rangle}+ A_{\langle a}A_{b\rangle}- 2\sigma_{c\langle a}\omega^c{}_{b\rangle}- \frac{10}{7}\,\sigma_{c\langle a}\sigma^c{}_{b\rangle}\,, \label{q2}\\ \overset{3}{\mathfrak{q}}_{abc}&=& -{\rm D}_{\langle a}\sigma_{bc\rangle}- 3A_{\langle a}\sigma_{bc\rangle} \hspace{15mm} {\rm and} \hspace{15mm}~ \overset{4}{\mathfrak{q}}_{abcd}=   \sigma_{\langle ab}\sigma_{cd\rangle}\,,  \label{q3-4}
\end{eqnarray}
are respectively the components of the associated monopole, dipole, quadruple, etc~\cite{H}. As before, these are also expressed in terms of the observers' irreducible kinematic variables.\footnote{Adopting the representation scheme of Eq.~(\ref{fancyQcomplete})-(\ref{q3-4}), decomposition (\ref{nH}) of the null Hubble parameter recasts as $\mathfrak{H}(e)=\overset{0}{\mathfrak{H}}+ \overset{1}{\mathfrak{H}}_ae^a+\overset{2}{\mathfrak{H}}_{ab}e^ae^b$, with $\overset{0}{\mathfrak{H}}=\Theta/3$, $\overset{1}{\mathfrak{H}}_a=-A_a$ and $\overset{2}{\mathfrak{H}}_{ab}=\sigma_{ab}$. Therefore dipolar anistropies in the null Hubble parameter are triggered by non-gravitational forces, while quadruple ones are due to shear distortions.} Following (\ref{q0})-(\ref{q3-4}), $\mathfrak{Q}$ and $q$ coincide in Friedmann universes, since $A_a=0=\sigma_{ab}= \omega_{ab}$ in FLRW models. Finally, we note that angled brackets indicate projected vectors and/or the projected symmetric traceless part of tensors (e.g.~$V_{\langle a\rangle}= h_a{}^bV_b$ and $T_{\langle ab\rangle}=h_{\langle a}{}^ch_{b\rangle}{}^d \;T_{cd}$ respectively).

\subsection{The case of an almost-FLRW universe}\label{ssCA-FLRWU}
The relations provided in \S~\ref{ssTDP} and in \S~\ref{ssNDP} are fully nonlinear and therefore hold in a general spacetime filled with any type of matter. As a result, these expressions can be linearised around any arbitrary background model of our choice. In what follows, we will assume that the aforementioned background spacetime is represented by the spatially homogeneous and isotropic FLRW cosmologies.

The high symmetry of the Friedmann models ensures that only time-dependent scalars survive, whereas vectors, tensors and spatial gradients vanish by default. Then, in a perturbed almost-FLRW universe, only the energy density ($\rho$) of the matter, its isotropic pressure ($p$), the expansion scalar ($\Theta=3H$) and the 3-Ricci scalar ($\mathcal{R}$ -- assuming there is spatial curvature) have nonzero background values. These are therefore the only variables of zero perturbative order. All the rest are linear (first-order) perturbations and therefore satisfy the Stewart~\&~Walker criterion for gauge invariance~\cite{StWa}.

Applying the linearisation scheme outlined above to decomposition (\ref{nH}), leaves the latter formally unchanged. This is not the case with expressions (\ref{fancyQcomplete}) and (\ref{q0})-(\ref{q3-4}), however, which simplify considerably and reduce to the first-order relations
\small
\begin{equation}
	\label{lfancyQcomplete}
\mathfrak{Q}(e)= -1- {1\over H^2} \left(\overset{0}{\mathfrak{q}}+{\overset{1}{\mathfrak{q}}}_ae^a +\overset{2}{\mathfrak{q}}_{ab}e^ae^b +{\overset{3}{\mathfrak{q}}}_{abc}e^ae^be^c +\overset{4}{\mathfrak{q}}_{abcd}e^ae^be^ce^d\right)\,,
\end{equation}
with
\begin{equation}
\overset{0}{\mathfrak{q}}= \frac{1}{3}\,\dot{\Theta}+ \frac{1}{3}\,{\rm D}_aA^a\,, \hspace{15mm} \overset{1}{\mathfrak{q}}_a= -\dot{A}_{\langle a\rangle}- \frac{1}{3}\,{\rm D}_a\Theta- \frac{2}{5}\,{\rm D}^b\sigma_{ab}\,, \label{lq0-1}
\end{equation}
\begin{equation}
\overset{2}{\mathfrak{q}}_{ab}= \dot{\sigma}_{\langle ab\rangle}+ {\rm D}_{\langle a}A_{b\rangle}\,, \hspace{15mm} \overset{3}{\mathfrak{q}}_{abc}= -{\rm D}_{\langle a}\sigma_{bc\rangle} \hspace{15mm} {\rm and} \hspace{15mm} \overset{4}{\mathfrak{q}}_{abcd}=   \sigma_{\langle ab}\sigma_{cd\rangle}\,.  \label{lq2-4}
\end{equation}
\normalsize
According to (\ref{lq0-1}a), in an almost-FLRW universe, the monopole component of the null deceleration parameter has an additional linear contribution due to the 4-acceleration and the non-gravitational forces it represents. The latter, together with spatial variations in the universal expansion and in the shear, also introduce dipole-like anisotropies in the $\mathfrak{Q}$-distribution (see Eq.~(\ref{lq0-1}b)). Finally, following expressions (\ref{lq2-4}), linear quadruple anisotropies in $\mathfrak{Q}$ are induced by the shear and the 4-acceleration, while the hihger order moments are solely due to the shear.

In what follows we will focus upon the monopole component ($\overset{0}{\mathfrak{q}}$) of the null deceleration parameter and try to estimate the changes introduced by the 4-acceleration term seen on the right-hand side of Eq.~(\ref{lq0-1}a). We will therefore ignore the dipole and all the higher order moments from now on. This means that, while in equation \eqref{lfancyQcomplete} $\mathfrak{Q}(e)$ represents the full null deceleration parameter, in what follows we will adopt the notation
\begin{equation}
\mathfrak{Q}= q- {1\over3H^2}\,{\rm D}_aA^a\,,  \label{lnQ}
\end{equation}
to represent its linear monopole component ($\overset{0}{\mathfrak{q}}$) in perturbed almost-FLRW universes. Assuming that matter is in the form of low-energy dust, the vanishing of the pressure guarantees zero 4-acceleration to linear order, which means that $\mathfrak{Q}=q$. This is no longer the case, however, when one allows for peculiar motions. Then, there is a non-zero (effective) 4-acceleration due to relative-motion effects alone (see~\cite{M} and also Eqs.~(\ref{A-tA}) here). Next, we will examine the implications of such an induced 4-acceleration for the null deceleration parameter.

\section{Tilted almost-FLRW universes}\label{sTA-FLRWUs}
Observations have repeatedly verified the existence of bulk peculiar flows (e.g.~see~\cite{ND}-\cite{Scetal}). Typically, these have sizes of few hundred Mpc and speeds of few hundred km/sec. Thus, typical galaxies like our Milky Way do not follow the (reference) CMB frame, but move relative to it.

\subsection{Two relatively moving frames}\label{ssTRMFs}
\begin{figure}[!tbp]
  \begin{subfigure}[b]{0.475\textwidth}
    \includegraphics[width=\textwidth]{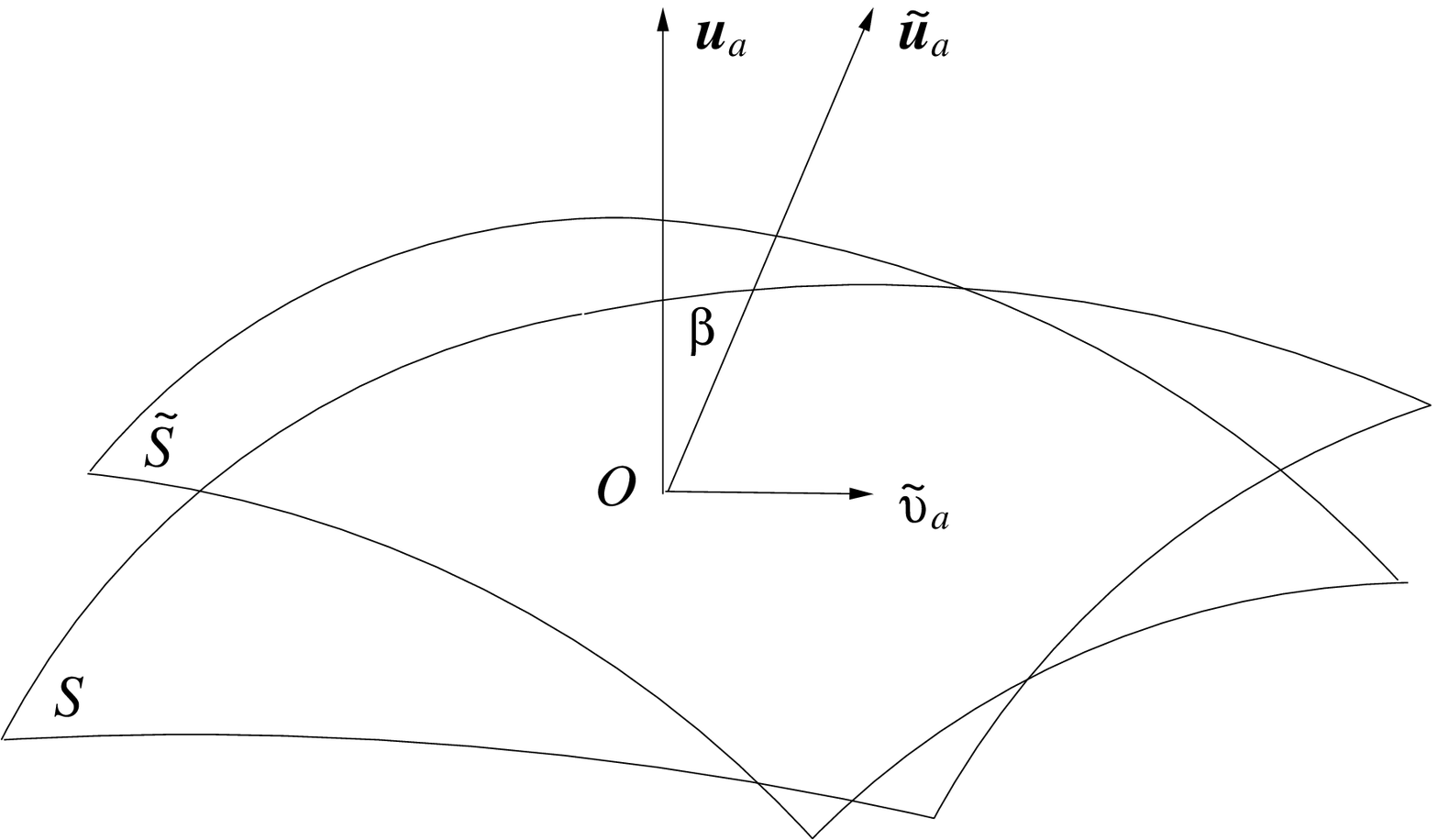}
    \caption{Tilted-frame perspective.}
    \label{fig:1}
  \end{subfigure}
  \hfill
  \begin{subfigure}[b]{0.475\textwidth}
    \includegraphics[width=\textwidth]{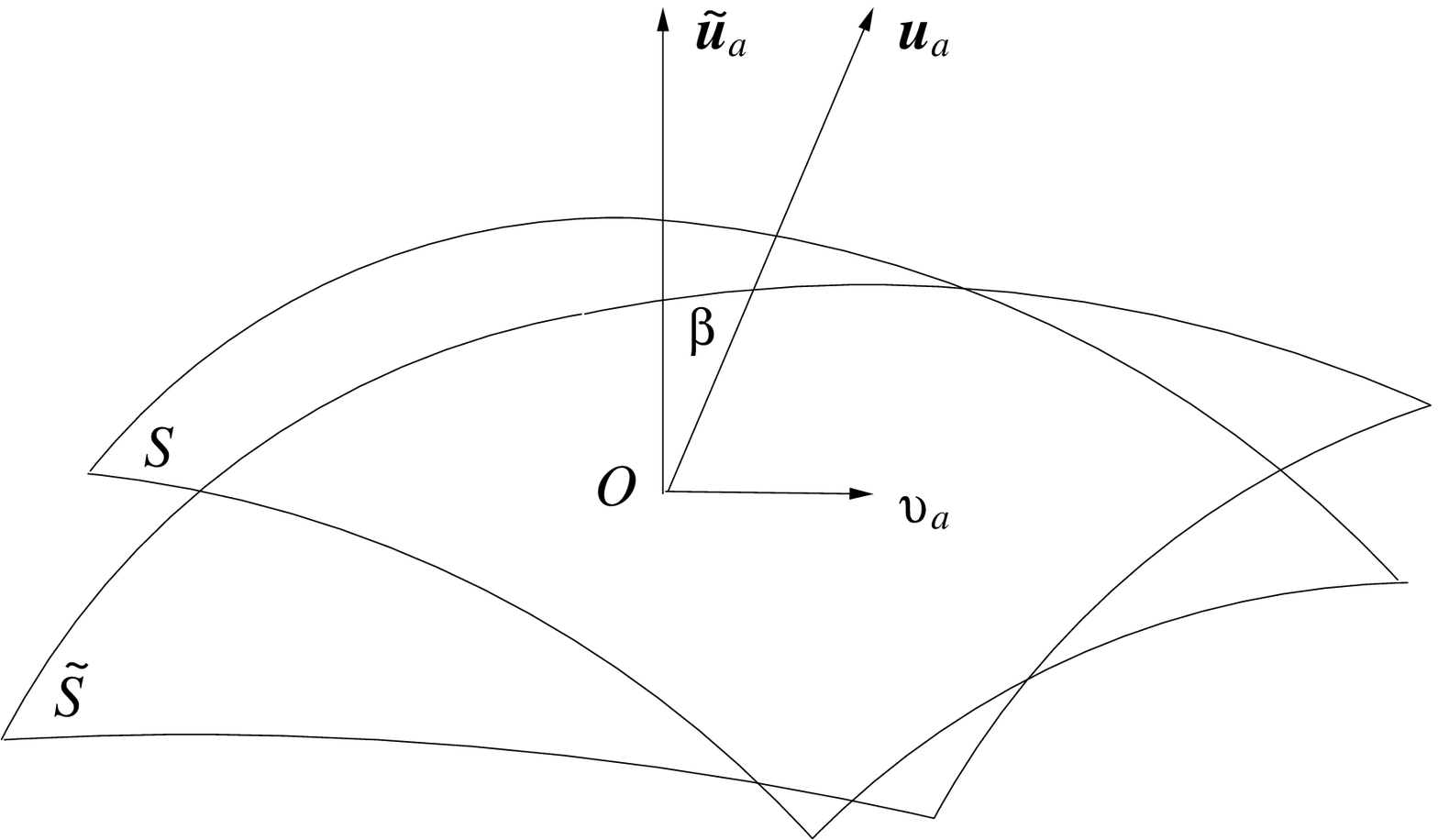}
    \caption{CMB-frame perspective.}
    \label{fig:2}
  \end{subfigure}
  \caption{Tilted spacetimes equipped with two families of observers, moving relative to each other with 4-velocities $u_a$ and $\tilde{u}_a$ at every event ($O$). Assuming that the $u_a$-field defines the reference coordinate system of the universe, $\tilde{u}_a$ is the 4-velocity of the real observers, ``drifting'' with peculiar velocity $\tilde{v}_a$ relative to the CMB-frame (see Eqs.~(\ref{4vels}a), (\ref{l4vels}a) and Fig.~\ref{fig:1} above). Alternatively, one may adopt the viewpoint of the idealised CMB observers. In that case, the latter are assumed to ``move'' relative to the real observers living in a typical galaxy with (effective) ``peculiar'' velocity $v_a$ (see Eqs.~(\ref{4vels}b), (\ref{l4vels}b) and Fig.~\ref{fig:2}). It goes without saying that both approaches lead to the same results.}  \label{fig:pmotion}
\end{figure}

In order to study the role and the implications of these bulk peculiar motions, one needs to employ cosmological models that allow for (at least) two families of observers, moving with respect to each other. By construction, one family is identified with the reference coordinate system of the universe, relative to which we can define and measure peculiar velocities. This is the idealised  CMB frame, where the associated dipole vanishes. The other frame corresponds to the real observers living in typical galaxies like ours and drifting relative to the CMB. In relativistic cosmology such models are known as ``tilted'', due to the (hyperbolic) tilt angle between the 4-velocities of the two observer groups (e.g.~see~\cite{KE,HUW} and also Fig.~\ref{fig:pmotion} here).

Let us therefore consider a perturbed FLRW universe and allow for two families of observers with (timelike) 4-velocities $\tilde{u}_a$ and $u_a$ respectively. Suppose also that the first group lives inside a bulk flow that moves relative to the CMB frame, which itself is identified with the second group. Then, the two 4-velocity fields are related by
\begin{equation}
\tilde{u}_a= \gamma(u_a+\tilde{v}_a) \hspace{15mm} {\rm and} \hspace{15mm} u_a= \gamma(\tilde{u}_a+v_a)\,,  \label{4vels}
\end{equation}
where $\gamma=(1-\tilde{v}^2)^{-1/2}=(1-v^2)^{-1/2}$ is the Lorentz-boost factor and $\tilde{v}^2=v^2$. Also, $\beta$ is the hyperbolic tilt angle between the two 4-velocities, with $\cosh\beta=-u_a\tilde{u}^a=\gamma\geq1$ (see Fig.~\ref{fig:pmotion}). For non-relativistic peculiar motions, we have $\tilde{v}^2,\,v^2\ll1$. Then $\gamma\simeq1$ and the above respectively reduce to
\begin{equation}
\tilde{u}_a= u_a+ \tilde{v}_a \hspace{15mm} {\rm and} \hspace{15mm} u_a= \tilde{u}_a+ v_a\,,  \label{l4vels}
\end{equation}
with $\tilde{v}_a=-v_a$ to first approximation.

When studying relative-motions, one is free to chose their reference coordinate system. In other words, one can either take the perspective of the (real) tilted observers (see Fig.~\ref{fig:pmotion}a), or that of their fictitious partners following the idealised CMB frame (see Fig.~\ref{fig:pmotion}b). Both approaches are equivalent and lead to the same results and conclusions, provided of course that all the effects have been properly accounted for. Thus, when taking the perspective of the tilted observers (see Fig.~\ref{fig:pmotion}a), one needs to remember that they move relative to their CMB counterparts with peculiar velocity $\tilde{v}_a$. When adopting the viewpoint of the idealised observers in the CMB frame, on the other hand, one needs to take into account that the latter ``move'' relative to the tilted frame with ``peculiar'' velocity $v_a$ (see Fig.~\ref{fig:pmotion}b). Once all the implications of relative motion have been accounted for, the two approaches are equivalent and lead to physically identical results. Although this should be expected, since physics is not frame dependent, we will verify our statement by taking both perspectives (see \S~\ref{ssSDRME} below).

Before closing, we should also point out that the above assume that there is a reference coordinate system in the universe (the CMB frame in our case), relative to which it makes sense to define and measure peculiar velocities. Although unlikely, it is conceivable that such a frame may not actually exist~\cite{Setal}-\cite{HPS}. In that case, the principles (at least) of the analysis presented here still hold, but now both 4-velocity fields correspond to real observers in distant galaxies moving relative to each other.

\subsection{$\mathfrak{Q}$ in the CMB and the tilted 
frames}\label{ssQTHFs}
Electromagnetic signals exchanged between observers in each of the aforementioned groups travel along null geodesics. Following~\cite{H}  -- see also Eq.~(\ref{lnQ}) here -- in an almost-Friedmann universe, the null and the timelike deceleration parameter associated with the CMB observers ($\mathfrak{Q}$) and the one measured by their bulk-flow counterparts ($\tilde{\mathfrak{Q}}$) are given by
\begin{equation}
\mathfrak{Q}= q- {1\over3H^2}\,{\rm D}^aA_a \hspace{15mm} {\rm and} \hspace{15mm} \tilde{\mathfrak{Q}}= \tilde{q}- {1\over3H^2}\,\tilde{{\rm D}}^a\tilde{A}_a\,,  \label{lQs1}
\end{equation}
to linear order.\footnote{Both groups of observers live in the real (the perturbed) universe, which means that (in addition to volume expansion) their 4-velocity fields have nonzero shear, vorticity and 4-acceleration as first-order perturbations (recall that the CMB is only nearly isotropic). Note, however, that neither the shear nor the vorticity are involved in our linear calculations, which makes these two tensors irrelevant for the purposes of this analysis (see also~\cite{T3,T4}).} Given that the CMB frame is the reference coordinate system of the cosmos, $q$ and $\mathfrak{Q}$ are the timelike and the null deceleration parameters of the whole universe, while $\tilde{q}$ and $\tilde{\mathfrak{Q}}$ are their local counterparts measured by the relatively moving (i.e.~the tilted) observers. Also note that $A_a=\dot{u}_a=u^b\nabla_bu_a$ and $\tilde{A}_a=\tilde{u}_a^{\prime}=\tilde{u}^b\nabla_b\tilde{u}_a$ are the 4-acceleration vectors in the aforementioned two coordinate systems, with $A_au^a=0=\tilde{A}_a\tilde{u}^a$ by construction. In addition, ${\rm D}_a= h_a{}^b\nabla_b$ and $\tilde{\rm D}_a=\tilde{h}_a{}^b\nabla_b$ are the spatial covariant derivative operators, relative to the $u_a$ and $\tilde{u}_a$ fields respectively.\footnote{Hereafter, tildas will be always associated with variables and operators defined in the bulk-flow frame. Also, primes will indicate time derivatives along the $\tilde{u}_a$-field, namely $\;{}^{\prime}=\tilde{u}^a\nabla_a$, while overdots will denote temporal differentiation in the $u_a$-frame (i.e.~$\;{}^{\cdot}=u^a\nabla_a$). Clearly, in the FLRW background the differential operators coincide.} Finally, $h_{ab}= g_{ab}+u_au_b$ and $\tilde{h}_{ab}=g_{ab}+\tilde{u}_a\tilde{u}_b$ are the associated projection tensors, with $g_{ab}$ being the spacetime metric and $h_{ab}u^b=0=\tilde{h}_{ab}\tilde{u}^b$ by default.

According to (\ref{lQs1}a) and (\ref{lQs1}b), the null deceleration parameters measured in the CMB and the bulk-flow observers are not necessarily equal, due to differences between the timelike deceleration parameters ($q$ and $\tilde{q}$) and between 4-acceleration vectors ($A_a$ and $\tilde{A}_a$) measured in the two frames. More specifically, we have
\begin{equation}
\mathfrak{Q}-\tilde{\mathfrak{Q}}= q- \tilde{q}- {1\over3H^2}\left({\rm D}^aA_a-\tilde{\rm D}^a\tilde{A}_a\right)\,.  \label{Q-tQ1}
\end{equation}
Next, we will evaluate the right-hand side of the above, assuming an FLRW background cosmology. In so doing, we will use the linear relations~\cite{M}
\begin{equation}
A_a- \tilde{A}_a= \dot{v}_a+Hv_a \hspace{15mm} {\rm and} \hspace{15mm} \tilde{A}_a- A_a= \tilde{v}_a^{\prime}+ H\tilde{v}_a\,,  \label{A-tA}
\end{equation}
with $\tilde{v}_a=-v_a$ and $\tilde{v}_a^{\prime}=-\dot{v}_a$ to first approximation (see also Fig.~\ref{fig:pmotion} in \S~\ref{ssTRMFs}).

Expressions (\ref{A-tA}) also serve as the linear propagation formulae of the peculiar velocity field. For instance, Eq.~(\ref{A-tA}a) recasts as $\dot{v}_a=-Hv_a+A_a-\tilde{A}_a$. Accordingly, peculiar velocities are generated, as well as sustained, by a nonzero difference between the 4-acceleration vectors and by the non-gravitational (in the relativistic sense) forces they represent. Indeed, setting $A_a-\tilde{A}_a$ to zero in Eq.~(\ref{A-tA}a) immediately leads to $\dot{v}_a= -Hv_a$, guaranteeing that peculiar velocities cannot be generated at the linear level. Moreover, any velocity perturbations that might have pre-existed will quickly decay away with the expansion (since $v\propto1/a$ on all scales). If so, given that peculiar velocities start very small at recombination, they should be all but eliminated by today. All this is clearly at odds with the observations and the established presence of bulk peculiar flows in the universe.

\section{Relative motion effects of $\mathfrak{Q}$}\label{sRMEQ}
Following (\ref{Q-tQ1}) and (\ref{A-tA}), the relative-motion effects on the null deceleration parameter are generally different from these on the timelike counterpart, namely $\mathfrak{Q}- \tilde{\mathfrak{Q}}\neq q-\tilde{q}$. In what follows, we will use relativistic linear cosmological perturbation theory to quantify this difference.

\subsection{Doubling the relative motion effect}\label{ssDRME}
Bulk peculiar flows are triggered by the ever increasing inhomogeneity and anisotropy of the post-recombination universe. Put another way, large-scale peculiar velocities are the inevitable byproduct of the ongoing process of structure formation. For instance, the Local Group and our Milky Way are drifting with respect to the (reference) CMB frame at approximately 600~km/sec, assuming that the CMB dipole is kinematic in nature.

Relative motions have been known to interfere with the way the associated observers interpret their data. In fact, the history of astronomy is rife with examples where relative-motion effects have lead to a gross misinterpretation of reality. Following~\cite{T1}-\cite{T3}, the measured value of the (timelike) deceleration parameter can be affected by the observers' motion with respect to the CMB frame. More specifically, in a tilted almost-FLRW universe, it was shown that
\begin{equation}
q- \tilde{q}= -{\dot{\vartheta}\over3H^2}- {2\vartheta\over3H}= {\tilde{\vartheta}^{\prime}\over3H^2}+ {2\tilde{\vartheta}\over3H}\,,  \label{q-tq1}
\end{equation}
where $\vartheta={\rm D}^av_a$ and $\tilde{\vartheta}=\tilde{\rm D}^a\tilde{v}_a$ (with $\tilde{\vartheta}=-\vartheta$ as well as $\tilde{\vartheta}^{\prime}= -\dot{\vartheta}$) at the linear level. Note that the scalar $\tilde{\vartheta}$ monitors the local expansion/contraction of the bulk flow in question, when positive/negative respectively. Taking the spatial divergence of (\ref{A-tA}a) and/or (\ref{A-tA}b) and keeping up to linear order terms, the above recasts as
\begin{equation}
q- \tilde{q}= -{1\over3H^2}\left({\rm D}^aA_a-\tilde{\rm D}^a\tilde{A}_a\right)\,,  \label{q-tq2}
\end{equation}
which substituted into Eq.~(\ref{Q-tQ1}) gives
\begin{equation}
\mathfrak{Q}- \tilde{\mathfrak{Q}}= -{2\over3H^2}\left({\rm D}^aA_a-\tilde{\rm D}^a\tilde{A}_a\right)= 2(q-\tilde{q})\,,  \label{Q-tQ2}
\end{equation}
Therefore, the linear relative-motion effect on the null deceleration parameter is twice stronger than on its timelike counterpart. In addition, this result applies to all FLRW backgrounds, irrespective of their spatial curvature and of the equation of state of the matter. Next, we will estimate the magnitude of the aforementioned effect, by taking a closer look at the role of the 4-acceleration terms seen on the right-hand side of the above.

\subsection{Scale dependence of the relative motion 
effect}\label{ssSDRME}
In line with (\ref{q-tq2}) and (\ref{Q-tQ2}), the linear effect of peculiar motions on both the timelike and the null deceleration parameters are triggered by differences between the 4-acceleration vectors associated with the CMB and the tilted observers. These represent the non-gravitational (in the relativistic sense) forces induced by structure-formation. The latter begins in earnest after recombination, when the universe is dominated by low energy dust (baryonic and/or not). In the absence of pressure, one can set the 4-acceleration to zero in either of the two frames, keeping in mind that it is nonzero in any other relatively moving coordinate system (see Eqs.~(\ref{A-tA}) in \S~\ref{ssQTHFs} and references therein).

Adopting the perspective of the idealised observers, namely those  following the CMB frame, we may set $\tilde{A}_a=0$ in the tilted coordinate system. Then, expression (\ref{Q-tQ2}) reduces to
\begin{equation}
\mathfrak{Q}- \tilde{\mathfrak{Q}}= -{2\over3H^2}\,{\rm D}^aA_a\,. \label{dQ-tQ1}
\end{equation}
The right-hand side term follows after applying linear relativistic cosmological perturbation theory to a tilted almost-FLRW universe with dust. In particular, linearising around the CMB frame the nonlinear Eq.~(2.3.1) of~\cite{TCM} (or equivalently Eq.~(10.101) of~\cite{EMM}), leads to
\begin{equation}
{2\over3H^2}\,{\rm D}^aA_a= {2\over3H^2}\left(\dot{\vartheta}+ 2H\vartheta\right)= {2\over9H^3}\,{\rm D}^2\vartheta- {2\over9}\,|1-\Omega|\left({\dot{\Delta}\over H} +{\mathcal{Z}\over H}\right)\,,  \label{DA1}
\end{equation}
where $\Omega$ is the density parameter of the Friedmann background. Also, $\Delta$ and $\tilde{Z}$ monitor scalar inhomogeneities in the density and the universal expansion respectively, as measured in the CMB frame. Given that our universe appears very close to spatial flatness (i.e.~$\Omega\rightarrow1$), the last term of the above is negligible.\footnote{We have assumed pressure-free matter and zero spatial curvature for mathematical simplicity. As shown in~\cite{T4}, the linear relative motion effects on the deceleration parameter persist in essentially all Friedman universes. It is also conceivable that the physical principles and the resulting effects discussed here go beyond the FLRW family of models.} Moreover, the spatial Laplacian on the right-hand side of (\ref{DA1}) implies a scale dependence, which becomes explicit after a simple harmonic decomposition. In so doing, expressions (\ref{dQ-tQ1}) and (\ref{DA1}) eventually combine to
\begin{equation}
\mathfrak{Q}- \tilde{\mathfrak{Q}}= {2\over9}\left({\lambda_H\over\lambda}\right)^2 {\vartheta\over H}\,. \label{dQ-tQ2}
\end{equation}
with $\lambda_H$ and $\lambda$ representing the Hubble horizon and the bulk-flow scale respectively~\cite{TK,T3}.

\begin{figure}[tbp]
\centering \vspace{5cm} \includegraphics{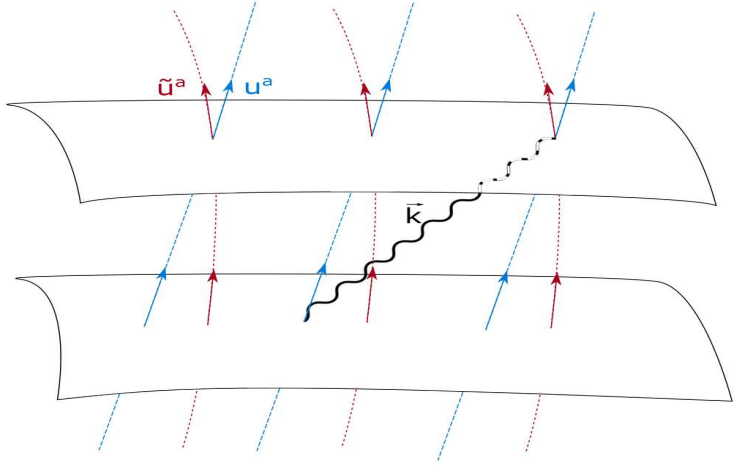} \caption{Radiation signals, travelling along the (null) $\vec{\rm k}$-direction, are received by the CMB and the (tilted) bulk-flow observers (associated with the blue and the red worldlines respectively). According to Eq.~(\ref{dtQ1}), the null deceleration parameters ($\mathfrak{Q}$ and $\tilde{\mathfrak{Q}}$) measured at the reception point by these two groups of observers differ due to relative-motions effects alone.}  \label{fig:null}
\end{figure}

Taking the viewpoint of the real observers, namely those living in typical galaxies like our Milky Way and moving with peculiar velocity $\tilde{v}_a$, we may set $A_a=0$ in the CMB frame. Then, Eq.~(\ref{Q-tQ2}) reads
\begin{equation}
\mathfrak{Q}- \tilde{\mathfrak{Q}}= {2\over3H^2}\,\tilde{\rm D}^a\tilde{A}_a\,. \label{dQ-tQ3}
\end{equation}
where now
\begin{equation}
{2\over3H^2}\,\tilde{\rm D}^a\tilde{A}_a= {2\over3H^2}\left(\tilde{\vartheta}^{\prime}+ 2H\tilde{\vartheta}\right)= {2\over9H^3}\,\tilde{\rm D}^2\tilde{\vartheta}- {2\over9}\,|1-\Omega|\left({{\tilde{\Delta}^{\prime}}\over H} +{\tilde{\mathcal{Z}}\over H}\right)\,,  \label{tDtA1}
\end{equation}
with $\tilde{\Delta}$ and $\tilde{\mathcal{Z}}$ being the ``tilted'' analogues of $\Delta$ and $\mathcal{Z}$ respectively. As before, the latter relation follows after linearising Eq.~(2.3.1) of~\cite{TCM}, or Eq.~(10.101) of~\cite{EMM}, this time in the tilted frame.\footnote{When deriving Eqs.~(\ref{DA1}) and (\ref{tDtA1}) we have also taken into account that the bulk peculiar flow induces an effective energy-flux vector, with $q_a=-\rho v_a$ and/or $\tilde{q}_a= -\rho\tilde{v}_a$ relative to the CMB and the tilted frames respectively, depending on the adopted perspective~\cite{M}. Substituting these linear relations into Eq.~(2.3.1) of~\cite{TCM} (or into Eq.~(10.101) of~\cite{EMM}) leads to expressions (\ref{DA1}) and (\ref{tDtA1}). Here, for the economy of the presentation, we have only provided the main results, referring the interested reader to~\cite{T3,T4} for the technical details and further discussion.} Again, the last term on the right-hand side of the above is negligible due to the near spatial flatness of our universe. Then, employing a simple harmonic splitting, we arrive at~\cite{TK,T3}.
\begin{equation}
\mathfrak{Q}- \tilde{\mathfrak{Q}}= -{2\over9}\left({\lambda_H\over\lambda}\right)^2 {\tilde{\vartheta}\over H}\,, \label{dQ-tQ4}
\end{equation}
which is identical to (\ref{dQ-tQ2}) since $\tilde{\vartheta}= -\vartheta$ to linear order. Consequently, in accord with (\ref{dQ-tQ2}) and (\ref{dQ-tQ4}), the null deceleration parameter measured by the real observers ($\tilde{\mathfrak{Q}}$) and the one measured by their idealised CMB counterparts ($\mathfrak{Q}$) (by means of light signals -- see Fig.~\ref{fig:null}) are related by
\begin{equation}
\tilde{\mathfrak{Q}}= \mathfrak{Q}+ {2\over9}\left({\lambda_H\over\lambda}\right)^2 {\tilde{\vartheta}\over H}\,, \label{dtQ1}
\end{equation}
with their difference being entirely due to relative motion effects. Moreover, the impact of the real observer's peculiar motion on $\mathfrak{Q}$ is twice stronger than on it timelike counterpart (recall that $\tilde{q}=q+ (1/9)(\lambda_H/\lambda)^2)(\vartheta/H)$ -- see~\cite{TK,T3} for details). Last, but not least, this result is independent of the adopted perspective. Put another way, relation (\ref{dtQ1}) holds both in the CMB and the tilted frames.

\subsection{Contracting bulk flows and the transition 
scale}\label{ssCBFTS}
Similarly to the timelike case, the effect of the observers peculiar motion is more sensitive to the scale-ratio ($\lambda_H/\lambda$), than to the velocity-ratio ($\tilde{\vartheta}/H$), which allows relatively slow bulk flows to have a disproportionately strong effect on scales well inside the Hubble radius. Also, as in the timelike case, the relative-motion effect depends on the sign of the local volume scalar ($\tilde{\vartheta}$), namely on whether the associated bulk flow is (locally) expanding or contracting (with $\tilde{\vartheta}\gtrless0$ respectively). Following (\ref{dtQ1}), observers inside expanding bulk peculiar motions measure $\tilde{\mathfrak{Q}}>\mathfrak{Q}$, while their counterparts in contracting bulk flows measure $\tilde{\mathfrak{Q}}<\mathfrak{Q}$. Moreover, the latter group of observers may also assign negative values to $\tilde{\mathfrak{Q}}$, entirely due to relative-motion effects. This happens on wavelengths smaller than the associated transition scale, which marks the $\tilde{\mathfrak{Q}}=0$ threshold and in our case it is given by
\begin{equation}
\lambda_\mathfrak{T}= \sqrt{{2\over9\mathfrak{Q}}\,{|\tilde{\vartheta}|\over H}}\, \lambda_H\,,  \label{tQlambdaT}
\end{equation}
in accord with Eq.~(\ref{dtQ1}).\footnote{The transition scale defined in (\ref{tQlambdaT}), as well as its timelike counterpart  ($q$), marks the wavelength below which the relative-motion effects dominate over the universal expansion, despite the weakness of the $\tilde{\vartheta}/H$-ratio. In this respect, $\lambda_\mathfrak{T}$ closely resembles the familiar ``Jeans length'' (see~\cite{T3} for a comparison and an extensive discussion). The latter, which also results from linear cosmological perturbation theory, sets the threshold below which weak pressure gradients dominate over the background gravity and dictate the linear evolution of density perturbations (e.g.~see~\cite{TCM,EMM}). In fact, the transition scale appears as generic to peculiar velocity perturbations as the Jeans length is to density perturbations~\cite{T3}.} Recalling that $\lambda_T=\sqrt{|\tilde{\vartheta}|/9qH}$ is the transition length for the timelike deceleration parameter (see~\cite{T3,T4} for details) and that $\mathfrak{Q}=q$ in the FLRW background (see Eqs.~(\ref{lQs1}) in \S~\ref{ssQTHFs} here), we deduce that $\lambda_{\mathfrak{T}}=\sqrt{2}\lambda_T$. In other words. the transition scale for $\mathfrak{Q}$ is approximately $\sqrt{2}\simeq1.4$ times larger than the one associated with its timelike counterpart. Finally, combining (\ref{dtQ1}) and (\ref{tQlambdaT}), we find that
\begin{equation}
\tilde{\mathfrak{Q}}^{\pm}= \mathfrak{Q}\left[1 \pm\left({\lambda_\mathfrak{T}\over\lambda}\right)^2\right]\,,  \label{ltQ-}
\end{equation}
where the plus/minus sign indicates (locally) expanding/contracting bulk flows. Therefore, focusing on contracting bulk peculiar motions, we have
\begin{equation}
\tilde{\mathfrak{Q}}^{-}\rightarrow \mathfrak{Q} \hspace{2mm} {\rm when} \hspace{2mm} \lambda\gg \lambda_\mathfrak{T}\,, \hspace{10mm} \tilde{\mathfrak{Q}}^{-}= 0 \hspace{2mm} {\rm at} \hspace{2mm} \lambda= \lambda_\mathfrak{T} \hspace{5mm} {\rm and} \hspace{5mm} \tilde{\mathfrak{Q}}^{-}< 0 \hspace{2mm} {\rm when} \hspace{2mm} \lambda<\lambda_\mathfrak{T}\,.  \label{ltQ-2}
\end{equation}
This is the theoretically predicted profile of the null deceleration parameter measured by observers inside a (locally) contracting bulk flow, in terms of scale (see also solid curve in Fig.~\ref{fig:Qplot} below). Not surprisingly, an exactly analogous profile has been reported for the timelike deceleration parameter (see Fig~2 in~\cite{T3,T4}). Note that the $\tilde{\mathfrak{Q}}$-profile seen above and depicted in Fig.~\ref{fig:Qplot}, as well as the numerical estimates given in Table~\ref{tab1} later, assume an average value of the local volume scalar (i.e.~we have set $\tilde{\vartheta}= \langle\tilde{\vartheta}\rangle$ -- see Table~\ref{tab1} here and also~\cite{T3}). Both the profile and the numerical estimates can be refined by introducing a scale dependence for $\tilde{\vartheta}$ (e.g.~see Fig.~3 in (\ref{tQlambdaT}) and discussion therein).

\begin{figure}[tbp]
\centering \includegraphics{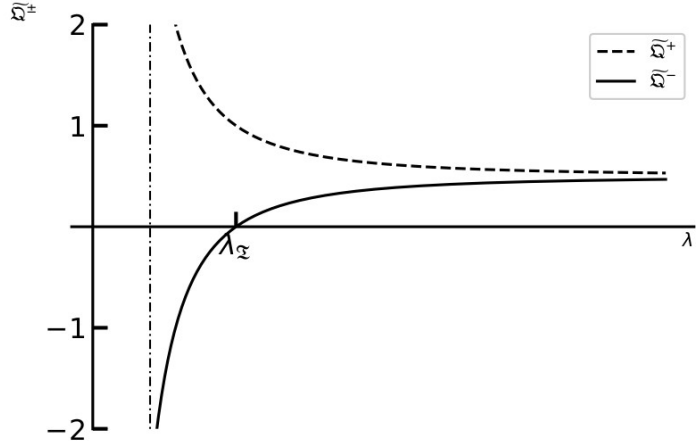} \vspace{5cm} \caption{The null transition length ($\lambda_\mathfrak{T}$) defined in (\ref{tQlambdaT}), assuming that $\mathfrak{Q}=1/2$ in the CMB frame. On scales much larger than $\lambda_\mathfrak{T}$ the relative-motion effects are negligible and $\tilde{\mathfrak{Q}}^{\,\pm}\rightarrow\mathfrak{Q}=1/2$. Scales inside the transition length, on the other hand, are ``contaminated'' by the observers peculiar motion. There, $\tilde{\mathfrak{Q}}^{\,+}$ becomes increasingly more positive (dashed curve), while $\tilde{\mathfrak{Q}}^{\,-}$ turns negative at $\lambda_\mathfrak{T}$ and keeps decreasing on progressively smaller wavelengths (solid curve). The vertical line marks the nonlinear cutoff, where the linear approximation is expected to break down. Note that the $\tilde{\mathfrak{Q}}$-profile depicted above can be refined by introducing a scale dependence for $\tilde{\vartheta}$ in (\ref{tQlambdaT}) -- see Fig.~3 in~\cite{AKPT}.}  \label{fig:Qplot}
\end{figure}

Overall, relative motions alone can locally assign (apparently) negative values to both the timelike and the null deceleration parameters (represented by $\tilde{q}^-$ and $\tilde{\mathfrak{Q}}^-$ respectively) in a universe that is globally decelerating (with $q,\,\mathfrak{Q}>0$ in the CMB frame). Thus, the accelerated expansion ``experienced'' by such observers is not real, but an artefact of their peculiar motion and more specifically of its local contraction. Since the relative-motion effects decrease with increasing scale, both $\tilde{q}^-$ and $\tilde{\mathfrak{Q}}^-$ become progressively less negative away from the observer. In fact, the two deceleration parameters cross the $\tilde{q}^-=0=\tilde{\mathfrak{Q}}^-$ mark at their corresponding transition lengths and start taking positive values beyond that threshold, eventually approaching their CMB-frame values (i.e.~$\tilde{q}^-\rightarrow q$ and $\tilde{\mathfrak{Q}}^-\rightarrow\mathfrak{Q}$) at sufficiently large scales/redshifts. Nevertheless, if the transition scale is large enough (e.g.~hundreds of Mpc --  see Table~\ref{tab1}), an unsuspecting observer may misinterpret the local change in the sign of the deceleration parameter as recent global acceleration.

Qualitatively speaking, the relative-motion effects on both the timelike and the null deceleration parameters are the same. Quantitatively, however the impact of the observers' peculiar flow is more pronounced on $\tilde{\mathfrak{Q}}$. The relative-motion effects can (locally) make $\tilde{\mathfrak{Q}}$ appear more negative than $\tilde{q}$. In addition, the transition scale for $\tilde{\mathfrak{Q}}$ is approximately 1.4 times larger than that for $\tilde{q}$ (see Table~\ref{tab1} for a numerical comparison of the relative-motion effects on $\tilde{q}$ and $\tilde{\mathcal{Q}}$ and on their transition scales).

\begin{table}
\caption{Representative estimates of the timelike and the null deceleration parameter ($\tilde{q}^-$ and $\tilde{\mathcal{Q}}^-$) and of their corresponding transition scales ($\lambda_T$ and $\lambda_\mathfrak{T}$), as measured in the rest-frame of the peculiar flows reported in~\cite{ND}-\cite{Scetal}. In all cases the bulk motions are assumed to (slightly) contract locally and the host universe is assumed to decelerate globally, with $q=\mathcal{Q}=0.5$ in the CMB frame. We have also set $H\simeq70$~km/sec\,Mpc and $\lambda_H= 1/H\simeq4\times10^3$~Mpc today. Finally, we have used the approximate relation $|\tilde{\vartheta}|/H\simeq\sqrt{3}\langle\tilde{v}\rangle/v_H$, where $\langle\tilde{v}\rangle$ is the reported mean bulk velocity on a certain scale and $v_H=\lambda H$ is the Hubble velocity on the same scale (see~\cite{T3} for more details).}
\begin{center}\begin{tabular}{ccccccccc}
\hline \hline & \hspace{-20pt} Survey & $\lambda$ (Mpc) & $\langle\tilde{v}\rangle$ (km/sec) & $\tilde{q}^-$ & $\lambda_T$ (Mpc) & $\tilde{\mathcal{Q}}^-$ & $\lambda_\mathfrak{T}$ (Mpc)& \\ \hline \hline & $\begin{array}{c} \hspace{-20pt} {\rm Nusser\,\&\,Davis} \\ \hspace{-20pt} {\rm Colin,\,et\,al} \\ \hspace{-20pt} {\rm Scrimgeour,\,et\,al} \\ \hspace{-20pt} {\rm Ma\,\&\,Pan} \end{array}$ & $\begin{array}{c} 280 \\ 250 \\ 200 \\ 170 \end{array}$ & $\begin{array}{c} 260 \\ 260 \\ 240 \\ 290 \end{array}$ & $\begin{array}{c} -0.01 \\ -0.24 \\ -0.81 \\ -2.05 \end{array}$ & $\begin{array}{c} 282 \\ 304 \\ 323 \\ 384 \end{array}$ & $\begin{array}{c} -0.51 \\ -0.97 \\ -2.11 \\ -4.60 \end{array}$ & $\begin{array}{c} 399 \\ 429 \\ 457 \\ 543 \end{array}$ \\ [2.5truemm] \hline \hline
\end{tabular}\end{center}  \label{tab1}
\end{table}

\section{Discussion}\label{sD}
Relative motions have been known to interfere with the way observers interpret their observations and understand their surroundings. Looking back into the history of astronomy, one can find a number of examples where such apparent effects have led to a serious misinterpretation of reality. \textit{Is it then conceivable that the recent universal acceleration may also be an illusion caused by our relative motion and, more specifically, by our peculiar flow with respect to the CMB expansion?}

This study, as well as analogous previous treatments, indicates that a positive answer to the above posed question is theoretically possible. Observers living in typical galaxies like our Milky Way, which drift relative to the CMB frame, can ``experience'' apparent local acceleration, within a globally decelerating universe, due to their peculiar motion alone. In particular, the deceleration parameters measured by the aforementioned observers can change sign, from positive to negative, entirely due to relative-motion effects. This occurs within locally contracting bulk peculiar flows and affects both the familiar timelike deceleration parameter ($q$), as well as its null counterpart ($\mathfrak{Q}$).\footnote{Assuming that there is no natural bias in favour of locally contracting or expanding bulk flows on scales of few hundred Mpc, the chances of an observer finding themselves living inside either of them should be roughly 50\%~\cite{T3}.} In both cases, the relative-motion effects become stronger closer to the observer (assumed -- for simplicity -- to reside at the centre of the bulk flow) and get weaker as one moves away form them. In both cases, the peculiar motion sets a characteristic length  scale, namely the \textit{transition length} ($\lambda_T$ and/or $\lambda_{\mathfrak{T}}$), where the locally measured deceleration parameters change sign. In either case, the host universe is still globally decelerating, since both $q$ and $\mathfrak{Q}$ remain positive in the reference (CMB) frame. Also, in either case, the unsuspecting bulk-flow observers are prone to misinterpret the (apparent) local change in the sign of their deceleration parameters as recent global acceleration. According to the present study, such a misinterpretation seems more likely to happen when ``measuring'' the null, rather than the timelike, deceleration parameter. The reason is that the effects of relative motion are more pronounced when acting on $\mathfrak{Q}$ rather than on $q$.

Evidence that our cosmological measurements might have been actually ``contaminated'' by relative-motion effects and that the recent universal acceleration could be a mere illusion, triggered by our peculiar motion with respect to the CMB frame, should be sought in the data. Reproducing the recent acceleration history of the universe to fit the observations is the first requirement for the viability of the tilted-universe scenario. In this respect, the profile $\tilde{\mathfrak{Q}}^-$ (as well as that of $\tilde{q}$), seen in Eq.~(\ref{ltQ-2}) and depicted in Fig.~\ref{fig:Qplot}, already reproduces the basic features of the deceleration parameter deduced by the observations. The aforementioned qualitative agreement was recently refined using the Pantheon supernovae dataset, the statistical analysis of which showed that the tilted model fits the data equally well with the $\Lambda$CDM paradigm (e.g.~see Fig.~3 in~\cite{AKPT}). Note that this has not been achieved by appealing to dark energy, to a cosmological constant, or after introducing any kind of new physics, but by simply accounting for the effects of the reported bulk peculiar motions

The data should also contain the ``trademark signature'' of relative motion, namely an apparent (Doppler-like) dipole anisotropy in the sky-distribution of the observed deceleration parameter, analogous to the dipole seen in the CMB spectrum. More specifically, both $\tilde{q}^-$ and $\tilde{\mathfrak{Q}}^-$ should take more negative values towards one direction in the sky and equally less negative in the opposite (see~\cite{T2,T3} for fairly thorough discussions). Put another way, to the bulk-flow observers, the universe should appear to accelerate faster along one direction on the celestial sphere and equally slower along the antipodal. In addition, assuming that the CMB dipole is also due to our local peculiar motion, the two dipolar axes should not lie far from each other. Over the last ten years or so, there has been a number of reports claiming that such a dipolar anisotropy actually exists in the supernovae data~\cite{SW}-\cite{BBA}. Nevertheless, it was only recently that the aforementioned dipole in the sky-distribution of the deceleration parameter was related/attributed to our peculiar motion relative to the CMB expansion~\cite{CMRS}. Additional support may also be coming from recent surveys claiming the presence of a dipole in the sky-distribution of the Hubble parameter as well~\cite{Metal1,Metal2}. Indeed, the very close connection between the Hubble and the deceleration parameters (definitions (\ref{H-q}) and (\ref{fancyh}) verify this), increases the possibility the two dipoles to be related. Finally, we note that the dipole in the sky-distribution of the deceleration parameter reported in~\cite{CMRS} appeared to make the associated monopole less negative, thus increasing the possibility the recent universal acceleration to be a mere relative-motion artefact. Future upcoming surveys and more refined data should help clarify whether such a theoretical scenario is observationally viable as well.

Speaking in more general terms, this study suggests that bulk peculiar flows are responsible for the formation of a region around every real observer in the universe, where the relative-motion effects dominate over the background expansion. There, observations can be heavily ``contaminated'' by relative-motion effects. Within the aforementioned region, the boundaries of which are set by the transition scale and vary between few hundred and several hundred Mpc, the local deceleration parameter (measured by the bulk-flow observers) and its global counterpart (measured by the idealised CMB observers) can vary considerably, both in their value and in their sign. This contamination seems to worsen when considering the null deceleration parameter associated with the (null) geodesics of the radiation signals, all of which make it imperative to account for the relative-motion effects before extrapolating conclusions drawn from locally collected data to the global universe.\\

\noindent\textbf{Acknowledgments:} The authors wish to thank Kerkyra Asvesta and Asta Heinesen for helpful discussions. This work was supported by the Hellenic Foundation for Research and Innovation (H.F.R.I.), under the ``First Call for H.F.R.I. Research Projects to support Faculty members and Researchers and the procurement of high-cost research equipment grant'' (Project Number: 789).

\end{document}